\documentclass[preprint,showpacs,preprintnumbers,amsmath,amssymb]{revtex4}

\usepackage{dcolumn}
\usepackage{bm}
\usepackage{epsfig}
\usepackage{hyperref}
\usepackage{graphicx}
\usepackage{dcolumn}
\usepackage{bm}
\usepackage{longtable}
\begin{document}
\title{Heavy-to-light transition form factors and their relations in light-cone QCD sum rules}
\author{Tao Huang$^{a,~b}$\footnote{Email: huangtao@mail.ihep.ac.cn.},
~Zuo-Hong Li$^c$\footnote{Email: lizh@ytu.edu.cn} and Fen
Zuo$^{a,~b}$\footnote{Email: zuof@mail.ihep.ac.cn.}}
\affiliation{
$^a$Institute of High Energy Physics, Chinese Academy
of Sciences,~Beijing 100049, China\\
$^b$Center for Science Facilities, Chinese Academy of Sciences,~Beijing 100049, China\\
$^c$Department of Physics, Yantai University, Yantai 264005,China}

\begin{abstract}
The improved light-cone QCD sum rules by using chiral current
correlator is systematically reviewed and applied to the calculation
of all the heavy-to-light form factors, including all the
semileptonic and penguin ones. By choosing suitable chiral currents,
the light-cone sum rules for all the form factors are greatly
simplified and depend mainly on one leading twist distribution
amplitude of the light meson. As a result, relations between these
form factors arise naturally. At the considered accuracy these
relations reproduce the results obtained in the literature.
Moreover, since the explicit dependence on the leading twist
distribution amplitudes is preserved, these relations may be more
useful to simulate the experimental data and extract the information
on the distribution amplitude.
\end{abstract}

 \maketitle

\section{Introduction}
Light-cone QCD sum rules have played an important role in the study
of heavy-to-light transitions. One of the main uncertainties of this
approach is due to the poorly known higher twist distribution
amplitudes of the final meson. To eliminate this uncertainty, Refs.
\cite{Chernyak1990,Huang2001} suggest to start from a correlator
composed of chiral currents in studying the form factor $f_+(q^2)$
of the $B\to \pi l\nu$ process. When chiral currents are used, the
opposite-parity state contributions add to the spectral density and
make it more smooth and more like the perturbative one. As a result,
the contributions from the twist-3 distribution amplitudes,
$\phi_p(u)$ and $\phi_\sigma(u)$, disappear from the resulting
light-cone sum rule. Moreover, it was found in Ref. \cite{Huang2001}
that all the twist-3 contributions, including that from the
three-particle Fock state, do not appear in this improved sum rule.
Thus up to twist-3 accuracy the sum rule for the form factor
$f_+(q^2)$ depends only on the leading twist distribution amplitude,
$\phi_\pi(u)$, so the result should be more stable \cite{Huang2001}.
Generally, one can prove that if chiral current is introduced, only
distribution amplitudes of the same chirality are maintained. For
the pseudoscalar meson, this immediately leads to the vanishing of
all the twist-3 terms, since all of them are of the opposite
chirality with the leading one. As a result, this improved sum rule
should be more suitable to determine the moments of $\phi_\pi(u)$
from the experimental results for $f_+(q^2)$ \cite{Wu2008}, compared
to ordinary sum rule \cite{Ball2006}. This improved sum rule was
further employed to study the semileptonic $B_s\to Kl\nu$
\cite{Huang2001b}, $B\to \eta l\nu$ \cite{Aliev2003}, and the
$B(B_c)\to Dl\nu$ \cite{Huang2006,Huang2007a} decays. This method
can be directly generalized to the calculation of the form factor
$f_-$ and the $B\to P$ penguin form factor $f_T$, leading to similar
sum rules as $f_+$. Since only the same distribution amplitude is
involved, simple relations between them can be easily found.

Except for the weak transitions of $B(B_s)$ into a pseudoscalar
meson, chiral currents were also utilized to calculate the radiative
form factor $T_1$ of the $B\to V\gamma$ process
\cite{Huang1998,Huang1999}, where $V$ denotes a light vector meson.
In this case, at the leading twist accuracy four distributions will
contribute. The dominant contribution comes from the leading twist
distribution amplitude of the transversely polarized vector meson,
$\phi_\perp(u)$, which is chiral-odd, while the others are all
chiral-even. As a result, when suitable chiral current is
introduced, we are left with a simplified sum rule depending only on
$\phi_\perp(u)$ at the leading twist accuracy. The numerical results
for the form factor and the corresponding branching ratio depend
crucially on the detail form of this distribution amplitude
\cite{Huang1999}. Combining with recent experiment result of ${\cal
B}(\bar B\to(\rho, \omega)+\gamma)$ \cite{Mohapatra2006}, this sum
rule can help to determine the properties corresponding distribution
amplitude. Generalization to the other penguin form factors and the
semileptonic form factors for the $B\to V$ transitions are also
straightforward.  Since all these sum rules depend mainly on one
distribution amplitude, simple relations arise naturally between
them.

 Literally, the relations for the heavy-to-light form factors
have been studied first by Stech \cite{Stech1995} and Soares
\cite{Soares1996} in the spectator quark model, and more explicitly
by Charles et. al \cite{Charles1999}, using light cone sum rules in
the limit of heavy quark mass for the initial hadron and large
energy for the final one. For the $B\to P$ transitions, our results
coincide with the relations obtained in Ref. \cite{Charles1999},
while in the $B\to V$ case our relations are nearly the same as the
leading power part of them. This is because in the later case only
the leading twist contributions have been considered in our
approach. Meanwhile, in our results the full dependence on the
distribution amplitudes has been maintained. Thus the relations in
our approach seems to be more general.

This paper is organized as follows. In the next section we review
the approach of using chiral currents in the light-cone QCD sum
rules, ad extend our previous results for $B\to Pl\nu$ and $B\to
V\gamma$ processes to all the heavy-to-light form factors. A
comparison of the relations between these form factors with other
approaches is given in Sec.{III}. The last section is reserved for
conclusion and discussion.

\section{The light-cone QCD sum rules with chiral currents}
\subsection{The $B\to P$ transition form factors}
Chiral current was first introduced into light-cone sum rule in Ref.
\cite{Chernyak1990}, in which the form factor $f_+(q^2)$ for
$B\to\pi l\nu$ at zero momentum was calculated. The chiral current
was also applied to the calculation of the B-meson decay constant
$f_B$ in ordinary QCD sum rule, leading to a suppression of power
corrections. A more explicit calculation of $f_+(q^2)$ for $B\to\pi
l\nu$ up to twist-4 terms in this approach, was given in Ref.
\cite{Huang2001}. Let us first review their strategy.

 ~~~The form factors $f_+(q^2)$ and $f_-(q^2)$ for the semileptonic
 $B\to Pl\nu$ transition are usually defined as:
\begin{equation}
\langle P (p)|\overline{u}\gamma _\mu b|B(p+q)\rangle=2f_+(q^2)
p_\mu +(f_+(q^2)+ f_-(q^2)) q_\mu \label{eq:BP}
\end{equation}
To obtain the relevant sum rules, one starts from the following
correlation function:
\begin{eqnarray}
F_\mu (p,q)&=&i\int d^4xe^{iqx}<P(p)\mid T\{\bar{q}(x)\gamma_\mu
b(x), \bar{b}(0)i\gamma_5 u(0)\}\mid 0>\nonumber\\
&=&F(p^2,(p+q)^2)p_\mu +\tilde{F}(p^2,(p+q)^2)q_\mu. \label{eq:1a}
\end{eqnarray}
The hadronic representation of this correlation function can be
obtained by inserting a complete set of states including the
$B$-meson ground state, higher resonances and non-resonant states
with B-meson quantum numbers:
\begin{eqnarray}
F_\mu (p,q)&=& \frac{<P\mid \overline q_2\gamma_\mu b\mid B> <B\mid
\bar{b}i\gamma_5q_1\mid 0>}{m_B^2-(p+q)^2}  \nonumber\\
&& +\sum_h \frac{ <P\mid \bar{s}\gamma_\mu b\mid h> <h\mid
\bar{b}i\gamma_5u\mid 0>}{m_h^2-(p+q)^2}\nonumber\\
&=&F(q^2,(p+q)^2)p_\mu +\tilde{F}(q^2,(p+q)^2)q_\mu \, .
\end{eqnarray}
Replacing the infinite sum by a general dispersion relation in the
momentum squared $(p+q)^2$ of the $B$-meson, one obtains:
\begin{equation}
F(q^2,(p+q)^2)= \int_{m_B^2}^\infty \frac{\rho (q^2,s)ds}{s-(p+q)^2}
\label{eq:HADRONRE}
\end{equation}
where possible subtractions are neglected and the spectral density
is given by
\begin{equation}
 \rho (q^2,s)=\delta
(s-m_B^2)2f_+(q^2)\frac{m_B^2f_B}{m_b}+ \rho^{h}(q^2,s) \,.
\label{4a}
\end{equation}
$\rho^h(p^2,s)$ denotes the spectral density of higher resonances
and of the continuum of states and can be replaced by
\begin{equation}
\rho^h(q^2,s)=\frac{1}{\pi}Im F_{QCD}(q^2,s)\Theta (s-s_0)
\label{6a}
\end{equation}
invoking quark-hadron duality. Here $s_0$ is the threshold
parameter, and $ImF_{QCD}(p^2,s)$ is obtained from the imaginary
part of the correlation function (\ref{eq:1a}) calculated in QCD.
This can be achieved by expanding the $T$- product of the current in
(\ref{eq:1a}) in the region of large space-like momenta
$(p+q)^2\ll0$. The leading contribution arises from the contraction
of the b-quark operators to the free $b$-quark propagator $ <0 \mid
b \bar{b}\mid 0> $ and involves the following distribution
amplitudes:
\begin{equation}
<P(p)\mid\bar{q_2}(x)\gamma_\mu\gamma_5q_1(0)\mid 0>= -ip_\mu f_P
\int_0^1due^{iupx} \varphi (u) \label{eq:pi2}
\end{equation}
\begin{equation}
<P(p)\mid \bar{q_2}(x)i\gamma_5q_1(0)\mid 0>= \frac{f_P
m_P^2}{m_1+m_2}\int_0^1due^{iupx}\varphi_p(u) \label{eq:pip}
\end{equation}

\begin{equation}
<P(p)\mid\bar{u}(x)\sigma_{\mu\nu}\gamma_5u(0)\mid 0>= i(p_\mu x_\nu
-p_\nu x_\mu )\frac{f_P m_P^2}{6(m_1+m_2)}
\int_0^1due^{iupx}\varphi_{\sigma }(u) \, . \label{eq:pisigma}
\end{equation}
where $m_1(m_2)$ is the current quark mass of $q_1(\overline q_2)$.
For the invariant amplitude $F$ the QCD representation reads:
\begin{eqnarray}
F_{QCD}(p^2,(p+q)^2) &=& -f_Pm_b\int_0^1du \frac{\varphi
(u)}{(q+up)^2-m_b^2} -\frac{f_P m_P^2}{m_1+m_2}\nonumber\\
&& \int_0^1du\left[\frac{\varphi_{p}(u)u}{(q+up)^2
-m_b^2}+\frac{\varphi_{\sigma } (u)}{6((q+up)^2-m_b^2)}
\left(2-\frac{p^2+m_b^2}{(q+up)^2-m_b^2}\right)\right] \,
.\nonumber\\
&&\label{eq:QCDRE}
\end{eqnarray}
Equating the Borel transformation of Eq. (\ref{eq:HADRONRE}) and
(\ref{eq:QCDRE}) we get the sum rule for the form factor
$f_+(q^2)$\cite{Belyaev1993}:
\begin{eqnarray}
f_+( q^2)&=& \frac{f_P
m_b^2}{2f_Bm_B^2}\int_{\Delta_P}^1\frac{du}{u}
\exp[\frac{m_B^2}{M_B^2}-\frac{m_b^2-\bar u(q^2-um_P^2)}{uM_B^2}] \nonumber\\
&&[ \varphi(u) + \frac{\mu_P}{m_b}u\varphi_{p}(u) +
\frac{\mu_P}{6m_b}\varphi_ {\sigma }(u)(2 +
\frac{m_b^2+q^2}{uM_B^2})] \, .\label{eq:fpi2}
\end{eqnarray}
where $\mu_P = m_P^2/(m_1+m_2)$ and $\Delta_P$ is the solution to
the equation $us_0^B-m_b^2-u\bar{u}m_P^2=0$ for $u\in[0,1]$.

In the above sum rule, the distribution amplitude $ \varphi(u)$ is
of twist-2, $\varphi_{p}(u)$ and $\varphi_ {\sigma }(u)$ are of
twist-3. There is also a twist-3 term from the following
three-particle operator:
\begin{eqnarray}
 <P(p)\mid
\bar{u}(x)gG_{\mu\nu}(z)\sigma_{\rho\lambda}\gamma_5u(0)\mid 0>
&=&if_{3P}[p_\mu (p_\rho g_{\lambda\nu}-p_\lambda
g_{\rho\nu})\nonumber\\
&&-q_\nu (q_\rho g_{\lambda\mu}-q_\lambda g_{\rho\mu})] \int{\cal
D}\alpha_i\varphi_{3K}(\alpha_i)e^{iq(x\alpha_1+z\alpha_3)}
\label{16a}
\end{eqnarray}
where $ G_{\mu\nu}(z)=(\lambda^c/2)G_{\mu\nu}^c(z)$, ${\cal
D}\alpha_i=d\alpha_1d\alpha_2d\alpha_3\delta
(\alpha_1+\alpha_2+\alpha_3-1)$, $\lambda^c$ and $ G_{\mu\nu}^c$
being the usual color matrices and the gluon field tensor. This term
comes from the b-quark propagator including the interaction with
gluons in first order:
\begin{eqnarray}
 <0\mid
T\{b(x)\bar{b}(0)\}\mid 0>_G&=&<0\mid T\{b(x)\bar{b}(0)\}\mid
0>-ig_s\int
\frac{d^4k}{(2\pi)^4}e^{-ikx}\int_0^1du\nonumber\\
&&\left[\frac{1}{2}\frac{k\!\!\!/+m_b}{(m_b^2-k^2)^2}G_{\mu\nu}(ux)\sigma^{\mu\nu}
+\frac{1}{m_b^2-k^2}ux-\mu G^{\mu\nu}(ux)\gamma_\nu\right]
\end{eqnarray}
Substituting this propagator into the original correlation function
and repeating the above process, we get the corresponding
corrections to the form factor $f_+(q^2)$ \cite{Belyaev1993}£º
\begin{eqnarray}
 &&f_{+G}(q^2)=-\frac{ f_{3P}m_b}{f_Bm_B^2 }\int_0^1u du\int
{\cal D}\alpha_i \Theta( \alpha_1+u\alpha_3-\Delta)\exp[\frac{m_B^2}{M_B^2}] \nonumber\\
&&\exp[-\frac{m_b^2-q^2(1-\alpha_1-u\alpha_3)}{(\alpha_1+
u\alpha_3)M_B^2}][1-\frac{ m^2_b -q^2 }{(\alpha_1+u\alpha_3)M_B^2}]
\frac{\varphi_{3\pi}(\alpha_i)}{(\alpha_1+ u\alpha_3 )^2}
\,.\nonumber\\
&&\label{eq:fpi3}
\end{eqnarray}
Eq.~(\ref{eq:fpi2}) and (\ref{eq:fpi3}) give the complete sum rule
for $f_+(q^2)$ at the accuracy of twist-3. However, the twist-3
distribution amplitudes are poorly known at present, which introduce
large uncertainties. To eliminate the twist-3 contributions, we can
start from the following correlation function with the $B$-meson
interpolating field $\overline{b}i\gamma_5q$ replaced by the chiral
current $\overline{b}i(1+\gamma _5)q$£º
\begin{eqnarray}
\Pi_\mu (p,q) && =i\int d^4xe^{iqx}\langle P (p)|T \{
\overline{q_2}(x)\gamma _\mu (1+\gamma
_5)b(x),\overline{b}(0)i(1+\gamma _5)q_1(0)\}|0\rangle \nonumber \\
&&=\Pi (q^2,(p+q)^2)p_\mu +\widetilde{\Pi }(q^2,(p+q)^2)q_\mu \,.
\end{eqnarray}
Now the scalar resonances corresponding to operator $\bar{b}q$,
which is of opposite parity to the $B$-meson, also add to the
spectral density. As a reflection of this fact, the QCD
representation of this correlation function contains only one single
distribution amplitude, at the accuracy of twist-3. In other words,
all the twist-3 contributions for this correlator disappear
automatically. More generally, one can prove that if chiral current
is introduced in the correlator, only the distribution amplitudes of
the same chirality remain in the final sum rule. In the pseudoscalar
case, one can see that all the twist-3 distributions are of opposite
chirality with the leading twist one, thus disappear automatically.
So up to twist-3 accuracy, we obtain the sum rule depending on
$\varphi(u)$ only:
\begin{equation}
f_+( q^2)=\frac{f_P m_b^2}{f_Bm_B^2}\int_\Delta^1\frac{du}{u}
\exp[\frac{m_B^2}{M_B^2}-\frac{m_b^2-\bar u(q^2-um_V^2)}{uM_B^2}]
\varphi(u)
\end{equation}
The sum rule for $f_-(q^2)$ can be obtained in the same way.
Actually, the QCD calculation of the corresponding correlation
function $\widetilde{\Pi }(q^2,(p+q)^2)$ vanished at the twist-3
accuracy, leading to the following relation
\begin{equation}
f_-(q^2)=-f_+(q^2).
\end{equation}

This method can be directly generalized to the calculation of the
penguin form factor $f_T(q^2)$, which is defined as:
\begin{equation}
\langle P ( p ) \left| \bar q  \sigma_{\mu\nu} q^\nu ( 1 + \gamma_5
) b \right| B (  p+q ) \rangle
 = i\frac{f_T(q^2)}{m_B+m_P}\left[ (2p+q)_\mu q^2-q_\mu(m_B^2-m_P^2)\right] ~,
\end{equation}
Starting from the standard correlation function
\begin{equation}
\Pi_\mu(p,q)=i\int d^4x~e^{ipx}
   \langle P(p)\vert T\{ \overline q_2 (x)i\sigma_{\mu\nu}q^\nu(1+\gamma_5)b(x)
  \, \bar b (0)i\gamma_5q_1(0)\} \vert  0 \rangle~,
\end{equation}
the corresponding sum rule has been derived in Ref.
\cite{Aliev1997}:
\begin{equation}
f_T(q^2)= \frac{m_b(m_B+m_P)f_P}{2f_Bm_B^2} \int_{\Delta_P}^1
\frac{du}{u}\exp[\frac{m_B^2}{M_B^2}-\frac{m_b^2-\bar
u(q^2-um_P^2)}{uM_B^2}]
\left[\varphi(u)+\frac{\mu_Pm_b}{3uM_B^2}\varphi_\sigma(u)\right]
\end{equation}
where the twist-4 terms has been omitted. As in the semileptonic
case, we simply replace the interpolating field $\bar b
i\gamma_5q_1$ by the left handed current $\bar{b}(0)
i(1-\gamma_5)q(0)$. Thus we start from the following correlation
function:
\begin{equation}
\Pi_\mu(p,q)=i\int d^4x~e^{ipx}
   \langle P(p)\vert T\{ \overline q_2 (x)i\sigma_{\mu\nu}q^\nu(1+\gamma_5)b(x)
  \, \bar b (0)i(1-\gamma_5)q_1(0)\} \vert  0 \rangle~.
\end{equation}
Repeating the procedure as in the previous case, the sum rule can be
obtained immediately:
\begin{equation}
f_T(q^2)= \frac{m_b(m_B+m_P)f_P}{f_Bm_B^2} \int_{\Delta_P}^1
\frac{du}{u}\exp[\frac{m_B^2}{M_B^2}-\frac{m_b^2-\bar
u(q^2-um_P^2)}{uM_B^2}] \varphi(u)
\end{equation}
Comparing with the sum rule for $f_+$, the following relation can be
easily found:
\begin{equation}
f_T(q^2)=\frac{m_B+m_P}{m_b}f_+(q^2)
\end{equation}
These relations between $f_+,~f_-$ and $f_T$ have been confirmed by
the numerical results in the light-cone QCD sum rules
\cite{Ball2005}. However, to make these relations manifest in the
ordinary light-cone sum rules, one needs to take certain limits, as
we will see in the next section.

\subsection{The $B\to V$ transition form factors}
In this subsection we will attempt to generalize the idea of chiral
current to the $B\to V$ transitions, where $V$ denotes a light
vector meson. At leading-twist accuracy, one will encounter the
following distributions: \cite{Ball1998}:
\begin{eqnarray}
\langle 0|\overline \psi_2(z) \gamma_{\mu}
\psi_1(-z)|V(P,\lambda)\rangle &=& f_{V} m_{V} \left[ p_{\mu}
\frac{e^{(\lambda)}\cdot z}{p \cdot z} \int_{0}^{1} \!du\, e^{i \xi
p \cdot z} \phi_{\parallel}(u) \right.
\nonumber \\
&&\left. + e^{(\lambda)}_{\perp \mu} \int_{0}^{1} \!du\, e^{i \xi p
\cdot z} g_{\perp}^{(v)}(u)\right] \label{eq:vda}
\end{eqnarray}
\begin{eqnarray}
\langle 0|\overline \psi_2(z) \gamma_{\mu} \gamma_{5}
\psi_1(-z)|V(P,\lambda)\rangle &=&
 \frac{1}{2}\left(f_V - f_V^{T}
\frac{m_{1} + m_{2}}{m_V}\right) m_V
\epsilon_{\mu}^{\phantom{\mu}\nu \alpha \beta} e^{(\lambda)}_{\perp
\nu} p_{\alpha} z_{\beta} \nonumber\\
&&\int_{0}^{1} \!du\, e^{i \xi p \cdot z} g^{(a)}_{\perp}(u).
\label{eq:avda} \\
\langle 0|\overline \psi_2(z) \sigma_{\mu \nu}
\psi_1(-z)|V(P,\lambda)\rangle &=& i f_{V}^{T} (
e^{(\lambda)}_{\perp \mu}p_\nu - e^{(\lambda)}_{\perp \nu}p_\mu )
\int_{0}^{1} \!du\, e^{i \xi p \cdot z} \phi_{\perp}(u).
\label{eq:tda}
\end{eqnarray}
where $\xi=2u-1,p_\mu=P_\mu-\frac{1}{2}z_\mu\frac{m_V^2}{pz}$. The
function $\phi_{\parallel}(u)$ and $\phi_{\perp}(u)$ give the
leading twist distributions in the fraction of total momentum
carried by the quark in transversely and longitudinal polarized
mesons, respectively. The functions $g_\perp^{(v)}$ ,
$g_\perp^{(a)}$ are always identified to be twist-3 from power
counting, but in fact they contain contributions of both operators
of twist-2 and twist-3 \cite{Braun1998}. Notice that
$\phi_{\perp}(u)$ is chiral-odd, while the other three are all
chiral-even. Therefore, by suitably choosing the chiral current one
may also obtain simplified sum rules at the leading-twist accuracy.
Let's demonstrate this procedure by reviewing the calculation of the
penguin form factor first. The relevant form factors are defined as
following:
\begin{eqnarray}
\langle V(p,\lambda) | \bar \psi \sigma_{\mu\nu} q^\nu (1+\gamma_5)
b | B(p_B)\rangle & = & i\epsilon_{\mu\nu\rho\sigma}
e^{(\lambda)\nu}
p_B^\rho p^\sigma \, 2 T_1(q^2)\nonumber\\
& & {} + T_2(q^2) \left\{ e^{(\lambda)}_\mu
  (m_B^2-m_{V}^2) - (e^{(\lambda)} p_B) \,(p_B+p)_\mu \right\}\nonumber\\
& & {} + T_3(q^2) (e^{(\lambda)} p_B) \left\{ q_\mu -
\frac{q^2}{m_B^2-m_{V}^2}\, (p_B+p)_\mu,
\right\}\nonumber\\
&&\label{eq:T}
\end{eqnarray}
where
\begin{equation}
 T_1(0)  =  T_2(0). \label{eq:T1T2}
\end{equation}
The decay width for the $B\to V\gamma$ process is mainly determined
by $T_1(0)$, so we can just focus on $T_1(q^2)$ only:
\begin{equation}
\langle V(p,\lambda) | \bar \psi \sigma_{\mu\nu} q^\nu \gamma_5 b |
B(p_B)\rangle  =  i\epsilon_{\mu\nu\rho\sigma} e^{(\lambda)\nu}
p_B^\rho p^\sigma \, 2 T_1(q^2).\label{eq:T1}
\end{equation}
To derive the light-cone sum rule for $T_1(q^2)$, usually one choose
the following correlation function based on Eq. (\ref{eq:T1}):
\begin{eqnarray}
T_{\mu}(p,q) &=&i\int d^{4}xe^{iqx}<V(p,\lambda)|T\overline
\psi_2(x) \sigma_{\mu\nu}\gamma_5 q^{\nu}b(x),\bar{b}(0) i\gamma_5
\psi_1(0)|0>\nonumber\\
&=&i\epsilon_{\mu\nu\rho\sigma} e^{(\lambda)\nu} p_B^\rho p^\sigma
\,T((p+q)^2)
\end{eqnarray}
A standard procedure leads to the following sum rule
\cite{Braun1994}:
\begin{eqnarray}
\lefteqn{\frac{f_B m_B^2}{m_b+m_q} 2 T_1(0) e^{-(m_B^2-m_b^2)/M_B^2}
=} \nonumber\\ &=&
 \int_0^1 du \frac{1}{u}\exp\left[-\frac{\bar u}{M_B^2}
\left(\frac{m_b^2}{u} + m_V^2\right)\right]
\theta\left[s_0-\frac{m_b^2}{u}-\bar u m_V^2\right] \Bigg\{m_b f^T_V
\phi_\perp(u,\mu)
\nonumber\\
&&\mbox{}+
 u m_V f_V g_\perp^{(v)}(u,\mu)
 +\frac{m_b^2-u^2m_V^2+u M_B^2}{ 4 u M_b^2} m_V f_V
 g_\perp^{(a)}(u,\mu) \Bigg\} \,.
\label{SR}
\end{eqnarray}
 Now we try to simplify the
sum rule by introducing suitable chiral current \cite{Huang1998}.
First, notice that Eq. (\ref{eq:T}) can be simplified when $q^2=0$:
\begin{eqnarray}
&&<V(p,\lambda)|\bar{\psi}\sigma_{\mu\nu}(1+\gamma_5)q^{\nu}b|B(p+q)>\nonumber\\
&=&2\{i\epsilon_{\mu\nu\alpha\beta}
                 e^{(\lambda)\nu}q^{\alpha}p^{\beta}
                 +p\cdot{q}e^{(\lambda)}_{\mu}
                 -q\cdot{e^{(\lambda)}}p_\mu \}T_1(0)+q\cdot{e^{(\lambda)}}q_\mu[T_3(0)-T_1(0)].\nonumber\\
&&
\end{eqnarray}
Starting from this definition, one can construct the following
correlator by choosing the right-handed current $\bar{b}
i(1+\gamma_5)q_1$ for the $B$ meson:
\begin{eqnarray}
F_{\mu}(p,q)=i\int d^{4}xe^{iqx}<V(p,\lambda)|T\overline \psi_2(x)
\sigma_{\mu\nu}
(1+\gamma_5) q^{\nu}b(x),\bar{b}(0) i(1+\gamma_5)\psi_1(0)|0>&&\nonumber\\
=\left[2i\epsilon_{
\mu\nu\alpha\beta}e^{(\lambda)\nu}q^{\alpha}p^{\beta}+2p\cdot
qe^{(\lambda)}_{\mu}-2q\cdot e^{(\lambda)}
p_{\mu}\right]F\left[(p+q)^2\right]+...~~~~~~~~&&\label{eq:HL}
\end{eqnarray}
Then the following simplified sum rule can be obtained:
\begin{eqnarray}
T_1(0)=\frac{m_b^2f^T_V}{m_{B}^2f_{B}}e^{m_{B}^2/M_B^2}
\int_{\Delta_V^0}^1{du\frac{\phi_{\perp}(u)}{u}\exp{[-\frac{m_b^2+u\bar{u}m_V^2)}{uM_B^2}]}},\label{eq:LCT1}
\end{eqnarray}
where $\Delta_V^0$ is the solution to the equation
$us_0-m_b^2-u\bar{u}m_V^2=0$ for $u\in[0,1]$.
So at the leading-twist accuracy we obtained a sum rule depending on
the distribution $\phi_{\perp}$ only, similar to the $B\to P$ case.
As a result, the final numerical results for $T_1(0)$ and the
branching ratio depend crucially on  $\phi_{\perp}$\cite{Huang1998}.
This fact can be utilized to determine the properties of
$\phi_{\perp}$ from the experimental results of the corresponding
decay process.

The sum rule for the form factor $T_1$ for finite value of $q^2$ can
be obtained from Eq. (\ref{eq:LCT1}) by trivial modifications. The
result is given as follows:
\begin{eqnarray}
T_1(q^2)&=&\frac{f^T_Vm_b^2}{f_Bm_B^2}e^{m_B^2/M^2_B}\nonumber\\
&&\times
\int^1_{\Delta_V}\frac{du}{u}\exp{\left[-\frac{m_b^2-\bar{u}(q^2-um_V^2)}{uM_B^2}\right]}\phi_\perp(u),
\end{eqnarray}
where
\begin{equation}
\Delta_V=[\sqrt{(s_0^B-q^2-m_V^2)^2+4m_V^2(m_b^2-q^2)}-(s_0^B-q^2-m_V^2)]/(2m_V^2),\label{eq:delta2}
\end{equation}
is the solution to $us_0^B-m_b^2-u\bar{u}m_V^2+\bar{u}q^2=0$~for
$u\in[0,1]$.  Generalization to the sum rules for other two form
factors $T_2$ and $T_3$ is also straightforward. First one can show
that the omitted terms of Eq. (\ref{eq:HL}) are identically zero at
the considered accuracy. Further decomposing Eq. (\ref{eq:HL}) in
the following form:
\begin{eqnarray}
F_{\mu}(p,q)&=&i\int d^{4}xe^{iqx}<V(p,\lambda)|T\overline \psi_2(x)
\sigma_{\mu\nu}
(1+\gamma_5) q^{\nu}b(x),\bar{b}(0) i(1+\gamma_5)\psi_1(0)|0>\nonumber\\
&=&\left[2i\epsilon_{
\mu\nu\alpha\beta}e^{(\lambda)\nu}q^{\alpha}p^{\beta}\right]F\left[(p+q)^2\right]\nonumber\\
&&+\left[ e^{(\lambda)}_\mu
  (m_B^2-m_{V}^2) - (e^{(\lambda)} p_B) \,(p_B+p)_\mu \right]\left(1-\frac{q^2}{m_B^2-m_{V}^2}\right)F\left[(p+q)^2\right]\nonumber\\
&&+\left[(e^{(\lambda)} p_B) q_\mu -
\frac{q^2}{m_B^2-m_{V}^2}\,(p_B+p)_\mu\right]F\left[(p+q)^2\right]\label{eq:HL2}
\end{eqnarray}
one immediately reads out the relations:
\begin{eqnarray}
T_2(q^2)&=&\left(1-\frac{q^2}{m_B^2-m_V^2}\right)T_1(q^2)\nonumber\\
T_3(q^2)&=&T_1(q^2)
\end{eqnarray}

Now we consider the semileptonic decay $B\to Vl\nu$. Generalization
of the chiral current method to this kind of process has been
attempted in Ref. \cite{Aliev2004}, where the interpolating field
for the heavy meson was chosen to be the left-handed chiral current.
The resulting sum rules contain the chiral-even terms of
$\phi_{\parallel},g^v_{\perp}$ and $g^a_{\perp}$, but the dominant
chiral-odd one, $\phi_{\perp}$, is eliminated. From the above
calculation for the penguin form factors,  it can be found that in
order to maintain the dominant contribution one should choose the
right-handed $\bar{b}(0) i(1+\gamma_5)q(0)$ instead. 
Let us specify this procedure more explicitly.

The form factors for the $B\to Vl\nu$ process can be defined as:
\begin{eqnarray}
&&\langle V(p,\lambda) | (V-A)_\mu | B \rangle  =  -i (m_B + m_V)
A_1(q^2) e^{(\lambda)}_\mu + \frac{iA_+(q^2)}{m_B
+ m_\rho} (e^{(\lambda)}p_B) (p_B+p)_\mu\nonumber\\
&& {} + \frac{iA_-(q^2)}{m_B + m_V} (e^{(\lambda)}p_B) (p_B-p)_\mu +
\frac{2V(q^2)}{m_B + m_V}
\epsilon_\mu^{\phantom{\mu}\alpha\beta\gamma}e^{(\lambda)}_\alpha
p_{B\beta} p_{\gamma},\makebox[0.8cm]{}\label{eq:ffV}
\end{eqnarray}
where $(V-A)_\mu = \overline \psi(z) \gamma_\mu(1-\gamma_5)b$ is the
corresponding weak current, $\lambda$ is the polarization vector of
the vector meson, and $q=p_B-p$ is the momentum transfer to the
leptons. Replacing the $B$ by the ordinary interpolating field
$j_B^\dagger = \overline b i\gamma_5 \psi_1$, one can consider the
following correlator:
\begin{eqnarray}
\Pi_\mu(p,q) & = & i\!\!\int d^4x\, e^{iqx}\, \langle V(p,\lambda) |
T (V-A)_\mu(z) j_B^\dagger (0) | 0 \rangle\nonumber\\
& = & -i \Pi_1((p+q)^2) e^{(\lambda)}_\mu + i\Pi_+((p+q)^2)
(e^{(\lambda)}p_B) (p_B+p)_\mu \nonumber\\
&&+ \Pi_V((p+q)^2)
\epsilon_\mu^{\phantom{\mu}\nu\alpha\beta}e^{(\lambda)}_\nu
p_{B\alpha} p_{\beta}+\dots,\label{eq:cor2pt}
\end{eqnarray}
where the term corresponding to the form factor $A_-$ is omitted for
simplicity. Repeating the procedure described in the previous
section, one obtain the following sum rules \cite{Ball1997}:
\begin{eqnarray}
\lefteqn{A_1(q^2)\ =\ \frac{m_b}{f_B(m_B+m_V)m_B^2}\, \exp\left\{
\frac{m_B^2-m_b^2}{M_B^2}\right\} \int_0^1
\frac{du}{u}\,\exp\left\{\frac{\bar
u}{uM_B^2}\,(q^2-m_b^2-um_V^2)\right\}}&&\nonumber\\
& & \Theta[c(u,s_0^B)]\left\{ f_\rho^\perp(\mu)\phi_\perp(u)\,
\frac{1}{2u}\,(m_b^2-q^2+u^2m_V^2) + f_V m_b m_V
g_\perp^{(v)}(u)\right\}\!,\nonumber\\
&&\label{eq:LCA1}\\
\lefteqn{A_+(q^2)\ =\ \frac{m_b(m_B+m_V)}{f_Bm_B^2}\,\exp \left\{
\frac{m_B^2-m_b^2}{M_B^2}\right\} \int_0^1
\frac{du}{u}\,\exp\left\{\frac{\bar
u}{uM_B^2}\,(q^2-m_b^2-um_V^2)\right\}}&&\nonumber\\
& & \left\{ \frac{1}{2}\,f_V^\perp(\mu) \phi_\perp(u)
\Theta[c(u,s_0^B)] + f_V m_b m_V \Phi_\parallel(u) \left[
\frac{1}{uM_B^2}\,
\Theta[c(u,s_0^B)] + \delta[c(u,s_0^B)]\right]\right\}\!,\nonumber\\
&&\label{eq:LCA2}\\
\lefteqn{V(q^2)\ =\ \frac{m_b(m_B+m_V)}{2f_Bm_B^2}\,\exp\left\{
\frac{m_B^2-m_b^2}{M_B^2}\right\} \int_0^1
\frac{du}{u}\,\exp\left\{\frac{\bar
u}{uM_B^2}\,(q^2-m_b^2-um_V^2)\right\}}&&\nonumber\\
& & \left\{ f_V^T \phi_\perp(u) \Theta[c(u,s_0^B)] +
\frac{1}{2}\,f_V m_b m_V g_\perp^{(a)}(u)\left[ \frac{1}{uM_B^2}\,
\Theta[c(u,s_0^B)] +
\delta[c(u,s_0^B)]\right]\right\}\!,\nonumber\\
&&\makebox[1cm]{\ } \label{eq:LCV}
\end{eqnarray}
where the definition
\begin{equation}
\Phi_\parallel(u,\mu) = \frac{1}{2}\,\left[ \bar u \int_0^u\!\!
dv\,\frac{\phi_\parallel(v,\mu)}{\bar v} - u \int_u^1\!\!
dv\,\frac{\phi_\parallel(v,\mu)}{v}\right]\label{eq:bigphi}
\end{equation}
has been used, and $c(u,s_0^B) = us_0^B-m_b^2+q^2\bar u-u\bar u
m_V^2$.

Now we replace the $j_B^\dagger = \overline b i\gamma_5 \psi$ in
Eq.(\ref{eq:cor2pt}) by the right-handed current $j_B^{R\dagger} =
\overline b i(1+\gamma_5) \psi_1$, and the corresponding correlator
becomes:
\begin{eqnarray}
\Pi_\mu(p,q)&=&-i\int d^4xe^{iqx}{<}V(p,\lambda)|T\{\overline \psi_2(x)\gamma_\mu(1-\gamma_5)b(x),\bar{b}_1(0)(1+\gamma_5)\psi_1(0)\}|0{>}\nonumber\\
            &=&\Gamma^1e^{(\lambda)}_\mu-\Gamma^+(e^{(\lambda)}q)(2p+q)q_\mu-\Gamma^-(e^{(\lambda)}q)q_\mu+i\Gamma^V\varepsilon_{\mu\alpha\beta\gamma}e^{(\lambda)\alpha}q^{\beta}p^{\gamma}.
\end{eqnarray}
A direct calculation leads to the following simplified sum rules:
\begin{eqnarray}
&&A_1(q^2)=\frac{f^T_Vm_b}{f_{B}m_B}e^{m_B^2/M^2_B}\nonumber\\
&&~~~~~~~\times \int^1_{\Delta_V}\frac{du}{u}\exp{\left[-\frac{m_b^2-\bar{u}(q^2-um_V^2)}{uM_B^2}\right]}\frac{m_b^2-q^2+u^2m_V^2}{um_B(m_B+m_V)}\phi_\perp(u),\label{eq:A1P}\\
&&\\
&&A_+(q^2)=\frac{f^T_Vm_b}{f_Bm_B}e^{m_B^2/M^2_B}\nonumber\\
&&~~~~~~~~~~~~\times \int^1_{\Delta_V}\frac{du}{u}\exp{\left[-\frac{m_b^2-\bar{u}(q^2-um_V^2)}{uM_B^2}\right]}\frac{(m_{B}+m_V)}{m_B}\phi_\perp(u),\\
&&A_-(q^2)=-A_+(q^2),\\
&&V(q^2)~~=A_+(q^2),
\end{eqnarray}
where
\begin{equation}
\Delta_V=[\sqrt{(s_0^B-q^2-m_V^2)^2+4m_V^2(m_b^2-q^2)}-(s_0^B-q^2-m_V^2)]/(2m_V^2),\label{eq:delta2}
\end{equation}
is the solution to $us_0^B-m_b^2-u\bar{u}m_V^2+\bar{u}q^2=0$~for
$u\in[0,1]$. Just as the improved sum rules for the penguin form
factors, these sum rules contain only the transverse distribution
amplitude $\phi_\perp(u)$, the chiral-even terms involving
$\phi_\parallel(u),~g^v_{\perp}$ and $g^a_{\perp}$ are completely
eliminated.

In Ref. \cite{Huang2006} and Ref. \cite{Huang2007b} We have
attempted to apply these sum rules in the $B\to Dl \nu$ process
 and the semileptonic decays of the
$B_c$-meson. The results for some channels, such as the $B\to Dl\nu$
and $B_c \to D(D^*)l\nu,B_c\to J/\psi(\eta_c)l\nu$, are roughly
consistent with other approaches. However, the best test background
for these sum rules should be the heavy-to-light transitions, so in
the following section we will compare our results with those derived
in other approaches, such as those in Ref. \cite{Stech1995} and Ref.
\cite{Charles1999}.

\section{Comparison with other approaches}
By using a constituent quark model approach and assuming simple
properties of the spectator quark, the semileptonic heavy-to-light
form factors are shown to be related by a single universal function
\cite{Stech1995}. In our definitions, these relations read:
\begin{eqnarray}
&&f_+(q^2)=R(q^2,m_F)\nonumber\\
&&f_-(q^2)=-R(q^2,m_F)\nonumber\\
&&A_1(q^2)=\frac{2E_F}{m_B+m_F}R(q^2,m_F)\nonumber\\
&&A_+(q^2,m_F)=\frac{m_B+m_F}{m_B}\
\frac{E_F-\frac{m^2_F}{m_B}}{m_F+E_F}R(q^2,m_F)\nonumber\\
&&A_-(q^2,m_F)=-\frac{m_B+m_F}{m_B}\
\frac{E_F+\frac{m^2_F}{m_B}}{m_F+E_F}R(q^2,m_F) \nonumber\\
&&V(q^2,m_F)=\frac{m_B+m_F}{m_B}R(q^2,m_F).\label{eq:Stech-Rela}
\end{eqnarray}
where $E_F=\frac{1} {2m_B}(m_B^2+m_F^2-q^2)$ is the energy of the
final state and $m_F$ denotes the mass. Furthermore, by employing
the Isgur-Wise relations\cite{Isgur1990} between the semileptonic
and the radiative form factors one can obtain \cite{Stech1995}
\begin{equation}
T_1(q^2)=R(q^2,m_F) \label{eq:IW1}
\end{equation}
These relations were further studied in Ref. \cite{Soares1996} and
\cite{Soares1998}.

Later, a more rigorous study of the form factor relations was done
in Ref. \cite{Charles1999}. Based on a light-cone sum rule
calculation in the limit of heavy mass for the initial hadron and
large energy for the final one, all the form factors are shown to
depend on three independent functions. Again we write the relations
in our present definition, which are as follows:
\begin{eqnarray}
f_+(q^2)&=&\zeta(m_B,E_P)\,,\label{eq:f+}\\
f_-(q^2)&=&-\zeta(m_B,E_P)\,,\label{eq:f-}\\
f_T(q^2)&=&\left(1+\frac{m_P}{m_B}\right)\zeta(m_B,E_P)\,,\label{eq:fT}\\
A_1(q^2)&=&\frac{2E_V}{M_B+m_V}\,\zeta_\perp(m_B,E_V)\,,\label{eq:A1}\\
A_+(q^2)&=&\left(1+\frac{m_V}{m_B}\right)\left[\zeta_\perp(m_B,E_V)-\frac{m_V}{E_V}\zeta_{/\!/}(m_B,E_V)\right]\,,\label{eq:A+}\\
A_-(q^2)&=&-\left(1+\frac{m_V}{m_B}\right)\left[\zeta_\perp(m_B,E_V)-\frac{m_V}{E_V}\zeta_{/\!/}(m_B,E_V)\right]\,,\label{eq:A-}\\
V(q^2)&=&\left(1+\frac{m_V}{m_B}\right)\zeta_\perp(m_B,E_V)\,,\label{V}\label{eq:V}\\
T_1(q^2)&=&\zeta_\perp(m_B,E_V)\,,\label{eq:T1}\\
T_2(q^2)&=&\left(1-\frac{q^2}{m_B^2-m_V^2}\right)\zeta_\perp(m_B,E_V)\,,\label{eq:T2}\\
T_3(q^2)&=&\zeta_\perp(m_B,E_V)-\frac{m_V}{E_V}\left(1-\frac{m_V^2}{m_B^2}\right)\zeta_{/\!/}(m_B,E_V)\,.
\label{eq:T3}
\end{eqnarray}
The three universal form factors $\zeta(M,E)$, $\zeta_{/\!/}(M,E)$
and $\zeta_\perp(M,E)$ are given by:
\begin{eqnarray}
\zeta(M,E)&=&\frac{1}{f_B}\frac{1}{2E^2}
\left[-f_P\phi^\prime(1)I_2(\omega_0,\mu_0)+\frac{f_Pm_P^2}{m_{q_1}+m_{q_2}}\phi_p(1)I_1(\omega_0,\mu_0)\right]\,,\label{eq:zeta1}\\
\zeta_{/\!/}(M,E)&=&\frac{1}{f_B}\frac{1}{2E^2}
\left[-f_V\phi_{/\!/}^\prime(1)I_2(\omega_0,\mu_0)+f_V^T m_Vh_{/\!/}^{(t)}(1)I_1(\omega_0,\mu_0)\right]\,,\label{eq:zeta2}\\
\zeta_\perp(M,E)&=&\frac{1}{f_B}\frac{1}{2E^2}
\left[-f_V^T\phi_\perp^\prime(1)I_2(\omega_0,\mu_0)+f_Vm_V
g_\perp^{(v)}(1)I_1(\omega_0,\mu_0)\right]\,.\label{eq:zeta3}
\end{eqnarray}
with the functions $I_j(\omega_0,\mu_0)$ defined by:
\begin{equation}
I_j(\omega_0,\mu_0)=\int_0^{\omega_0}d\omega\,\omega^j\exp\left[\frac{2}{\mu_0}\left(
\overline{\Lambda}-\omega\right)\right]\ \ \ \ \ j=1,\,2
\end{equation}
Here the parameters $\overline{\Lambda},~\mu_0$ and $\omega_0$ are
related to the ordinary parameters:
\begin{eqnarray}
\overline{\Lambda}&=&m_B-m_b\nonumber\\
M_B^2&=&m_b\mu_0\nonumber\\
s^B_0&=&(m_b+\omega_0)^2\,,
\end{eqnarray}
These relations are confirmed in the Soft-Collinear Effective Theory
\cite{Bauer2001}. The above relations (\ref{eq:f+})-(\ref{eq:T3})
are quite similar to those (Eq. (\ref{eq:Stech-Rela})  and Eq.
(\ref{eq:IW1})) obtained by Stech. Actually, if one impose
$\zeta=\zeta_\perp=\zeta_{/\!/}$, these two set of relation almost
coincide except some ambiguity in the sub-leading terms $\sim
m_F^\prime/m_B$ or $m_F^\prime/E_F$ \cite{Charles1999}. Although
from the expressions (\ref{eq:zeta1})-(\ref{eq:zeta3}) one can not
find general reasons for this relation to hold, the numerical
results for these three form factors may support it, because the
decay constant and the leading twist distribution amplitudes of the
pseudoscalar, the longitudinally and the transversely polarized
vector meson are not quite different. This explains in some sense
the consistence of the relations obtained by Stech with the lattice
data \cite{Debbio1998}.


Now compare our results with those given in Eqs.
(\ref{eq:f+})-(\ref{eq:T3}). For the $B\to P$ transitions form
factors the relations from the two approaches are exactly the same.
As have been mentioned in previous section, these relations were
also confirmed by the numerical results in the light-cone sum rule
calculation \cite{Ball2005}. For the $B\to V$ transitions, our
leading-twist results are also very similar as the leading power
part of Eqs. (\ref{eq:A1})-(\ref{eq:T3}). The only difference is in
the extra factor $2E_V/(m_B+m_V)$ for $A_1$. In our result this
factor is $u$-dependent and inside the integral over $u$
(\ref{eq:A1P}). However, when $E_V$ is taken to be very large,
$\Delta_V\to 1$ this factor
$\frac{m_b^2-q^2+u^2m_V^2}{um_B(m_B+m_V)}\sim \frac{2E_V}{m_B+m_V}$
and factors out. Thus at the considered accuracy, our approach by
using the chiral currents reproduces naturally the corresponding
relations obtained from other approaches, and at the same time
preserves the full dependence on the leading twist distribution
amplitudes. So our relations can be directly utilized to simulate
the experimental data and extract the corresponding information on
the distribution amplitudes.

\section{Conclusion}
The improving approach of using chiral currents in the light-cone
QCD sum rules is systematic reviewed and successfully generalized to
all heavy-to-light weak transition. The resulting light-cone sum
rules for all the semileptonic and penguin form factors depend only
on one leading twist distribution amplitude, up to twist-3 accuracy
for the $B\to P$ transitions and to leading-twist accuracy in the
$B\to V$ case. The other contributions disappear automatically since
they have the opposite chirality with the dominant one. Since the
poorly-known twist-3 distribution amplitudes are eliminated, these
sum rules should be more stable than the ordinary one. A systematic
numerical calculation of the heavy-to-light form factors using these
sum rules is in process. Moreover, if the form factors is known very
well experimentally, one can also utilize these sum rules to study
the properties of the leading twist distribution
amplitudes~\cite{Wu2008}.

Since only one leading distribution amplitude is involved, simple
relations for all the form factors arise naturally in our approach.
At the considered accuracy these relations reproduce the results
obtained by using light-cone sum rules in the limit of heavy quark
mass for the initial hadron and large energy for the final one, and
at the same time preserve the full dependence on the leading twist
distribution amplitude. Therefore these relations may be more useful
to simulate the experimental data and extract the information of the
leading twist distribution amplitudes.

\begin{center}
{\bf ACKNOWLEDGEMENTS}
\end{center}

This work was supported in part by the Natural Science Foundation
of China (NSFC) under Grant No 10475084.  \\

\newpage

\end{document}